\documentclass[aps,prl,reprint,onecolumn,superscriptaddress,longbibliography,notitlepage]{revtex4-1}

\usepackage{graphicx}
\usepackage{rotating}
\usepackage{color}
\usepackage{soul}
\usepackage[normalem]{ulem}
\usepackage{caption}
\usepackage{subcaption}
\usepackage{amsmath,amssymb}
\usepackage[english]{babel}
\usepackage{natbib}
\usepackage{upgreek}
 
\begin{document}
\title{Optical trapping and critical Casimir forces}
\date{Received: date / Revised version: date}

%
\author{Agnese Callegari}
\affiliation{Department of Physics, University of Gothenburg, 41296 Gothenburg, Sweden.}
\author{Alessandro Magazz\`u}
\email{alessandro.magazzu@physics.gu.se}
\affiliation{Department of Physics, University of Gothenburg, 41296 Gothenburg, Sweden.}
\author{Andrea Gambassi}
\affiliation{SISSA -- €"International School for Advanced Studies, 34136 Trieste, Italy.}
\altaffiliation[Also at ]{INFN, Sezione di Trieste, 34136 Trieste, Italy.}
\author{Giovanni Volpe}
\affiliation{Department of Physics, University of Gothenburg, 41296 Gothenburg, Sweden.}

\begin{abstract}
Critical Casimir forces emerge between objects, such as colloidal particles, whenever their surfaces spatially confine the fluctuations of the order parameter of a critical liquid used as a solvent. These forces act at short but microscopically large distances between these objects, reaching often hundreds of nanometers. Keeping colloids at such distances is a major experimental challenge, which can be addressed by the means of optical tweezers. 
Here, we review how optical tweezers have been successfully used to quantitatively study critical Casimir forces acting on particles in suspensions. 
As we will see, the use of optical tweezers to experimentally study critical Casimir forces can play a crucial role in developing nano-technologies, representing an innovative way to realize self-assembled devices at the nano- and microscale.
\end{abstract}

\maketitle

\section{Introduction}\label{sec:1:intro}

In 1948, H. Casimir predicted the existence of an attractive force between two identical, parallel, uncharged, metallic plates placed in vacuum, at zero temperature and at a short distance from each other\cite{casimir1948attraction}. This force is of pure quantum electro-dynamical (QED) nature: the electromagnetic field between the plates is allowed to assume only certain frequencies, because it must satisfy the boundary conditions imposed by the presence of the conducting surfaces. The force between the plates turns out to be attractive, and, upon increasing the distance $d$ between the plates, decays as $d^{-4}$. A first attempt at measuring QED Casimir attraction was performed in 1958 by M. Spaarnay at Philips in Eindhoven but was not conclusive, because of its large experimental uncertainties. A series of more successful measurements begun in 1997:  Lamoreux measured the Casimir force between a flat plate and a spherical conducting surface with a large radius using a torsion pendulum\cite{lamoreaux1997}, and in 2002 the force between parallel metallic surfaces was measured in an experiment with microresonators\cite{bressi2002}.
The key ingredient for the emergence of the Casimir attraction is the presence of a fluctuating field (the quantum electromagnetic field) spatially confined between two objects (the conducting plates).

Casimir-like forces can be observed in contexts other than QED, whenever a fluctuating field is spatially confined by the presence of objects. For example, a macroscopic, maritime analogue of the Casimir effect is the attraction observed between parallel boats located at a short distance from each other\cite{boersma1996}. In 1978, Fisher and de Gennes \cite{fisher1978phenomena} showed that Casimir-like forces may arise also in soft matter systems. They considered the case of a critical binary liquid mixture, i.e., a mixture of two different components characterized by a second-order phase transition at a critical temperature $T_{\rm c}$ and critical concentration $\chi_{\rm c}$. When the temperature $T$ approaches the critical temperature at the critical concentration, the spatial correlation length of the fluctuations of the local concentration of one of the two components with respect to its critical value increases, and ideally it diverges when the critical temperature is reached. If the liquid mixture is spatially confined by walls, a force arises on them, with a spatial range set by the correlation length of the critical fluctuations. This force has been termed the {\em critical} Casimir force (CCF), because it emerges in confined fluids close to criticality. 
The concentration fluctuations in the critical mixture play a role similar to that played by the quantum fluctuation of the electromagnetic field confined between the two conducting plates \cite{gambassi2009casimir,gambassi2011critical}.

The first indirect experimental confirmation of the occurrence of the critical Casimir effect was provided by a set of works concerning the reversible flocculation of colloids in binary liquid mixtures \cite{Beysens1985,Narayanan1993}, where attractive forces acting on the colloids suspended in a near-critical binary liquid mixture lead to aggregation upon approaching the critical temperature.
It is worth noting that the results of these works were not originally explained in terms of the occurrence of CCFs but rather in terms of other phenomena, closely related to wetting, i.e.,  capillary condensation due preferential adsorption and bridging. In a retrospective examination, however, part of the described phenomena turns out to be consistent with the action of CCFs.

Critical Casimir forces started to be investigated experimentally on wetting films of fluids close to criticality in the late 1990s and early 2000s \cite{Garcia1999,Fukuto2005,Ganshin2006}. A wetting  film, in fact, provides an experimental realization for the geometry of two parallel planar walls confining a liquid phase. In fact, the wetting film of a liquid phase is actually confined between its interface with the solid substrate at which the wetting film forms and the interface with its vapor phase. The fluctuations of the order parameter of a possible second-order phase transition occurring within the liquid phase are therefore confined within the film of a certain thickness $L$. Upon approaching the critical temperature of this phase transition, a critical Casimir pressure emerges on the two surfaces and its action on the liquid-vapour interface causes the film thickness to change from its equilibrium value in the absence of fluctuations. This change depends on the various thermodynamic control parameters of the system and in particular on the temperature: knowing the relation between thickness and pressure, the magnitude of the CCF can therefore be inferred.
The early measurement of CCFs via wetting films of $^{4}$He or $^{4}$He-$^{3}$He mixtures near the normal-superfluid transition \cite{Garcia1999,garcia2000,Ganshin2006} and of  binary mixtures at their demixing transition \cite{Fukuto2005,rafai2007}
were based on this approach, and provided the first indirect determination of CCFs. 
These and other early experimental investigations relied primarily on using  X-rays\cite{Fukuto2005} or light scattering\cite{Beysens1985,Narayanan1993}, capacitance measurement\cite{Garcia1999,garcia2000,Ganshin2006}, as well as ellipsometry\cite{mukhopadhyay1999,rafai2007}.

Only in 2008 were CCFs measured directly via total internal reflection microscopy (TIRM) \cite{hertlein2008direct,gambassi2009critical}, by direct reconstruction of their potential in the case of a spherical particle near a planar surface.
In particular, the authors of Refs.~\cite{hertlein2008direct,gambassi2009critical} measured the CCFs arising between an optically trapped colloidal particle and a planar surface, immersed in a binary liquid mixture of water and 2,6-lutidine, at several temperatures approaching its lower critical point at $T_{\rm c} \simeq 34^\circ {\rm C}$. They measured the statistics of the distance of the particle from the surface by TIRM \cite{prieve1999measurement, volpe2009novel} as function of the temperature, which allowed them to determine the corresponding potential of the forces acting on the particle, with femtonewton resolution  \cite{hertlein2008direct}. In addition, depending on the gross chemical properties of the spherical surface of the colloid and of the planar surface, i.e., them being hydrophilic/hydrophobic, they were able to measure experimentally attractive or repulsive forces, in striking quantitative agreement with the theoretical predictions for the CCF interaction, showing an exquisite temperature dependence of these forces \cite{gambassi2009critical}.

Experimentally, the critical Casimir potential is often obtained from observing and determining the trajectory of one of the objects (usually a particle) subject to the CCF and to possible additional forces, diffusing within the liquid solvent. 
In practice, the particle, due to its Brownian motion within the medium, explores a region of space corresponding to various distances from another neighbouring object (particle, surface). In equilibrium, these distances are sampled according to the Boltzmann distribution involving the potential of the forces acting on the particle, which can therefore be determined from the distribution of the experimental data, together with the corresponding total force acting on the fluctuating particle.

Optical tweezers\cite{ashkin1970acceleration,ashkin1986observation,jones2015optical,gieseler2020optical} are versatile tools which allow one to trap and manipulate micro and nano-particles in liquids with nanometric precision as well as to measure forces with femtonewton resolution \cite{marago2013optical, irrera2016photonic,marago2010photonic}.
The classical optical tweezers is realized by focusing a laser beam through an high numerical aperture objective. When a particles is in the proximity of the waist of the focused beam, it
scatters the light hitting its surface and the momentum exchanged with light gives rise to an optical potential which is harmonic for small displacement of the particle from its equilibrium position \cite{ashkin1986observation,jones2015optical,gieseler2020optical}. Thus, a linear restoring force pulls the particle back towards the center of the optical trap and one can quantify forces acting on the particles, e.g., by measuring the particle displacement from its equilibrium position\cite{jones2015optical,gieseler2020optical}.

By shaping the phase of a laser beam with a spatial light modulator before focusing it through the high numerical aperture objective, it is possible to generate multiple reconfigurable optical traps.
By tuning the stiffnesses of these optical traps, it is possible to let the particles  fluctuate and explore the surrounding of their equilibrium positions and therefore to {\em map} the total potential in their proximity, which consists of an external contribution (due to, e.g., optical and gravitational forces) and an interparticle interaction conribution, which may include  electrostatic, critical Casimir, and van der Waals interactions. Accordingly, optical tweezers can play a crucial role in the investigation of CCFs between multiple particles including the emergence of many-body effects. \cite{paladugu2016nonadditivity, magazzu2019controlling}.

Although for several years CCFs (as well as the QED Casimir effect mentioned at the beginning) were out of reach for experiments and considered to have limited potential applications, the situation has changed in the last two decades: in fact, they were recently employed for manipulating and assembling structures and colloidal molecules \cite{bonn2009direct,marino2016assembling,gambassi2010colloidal,faber2013controlling,nguyen2017tuning}.
A proof of concept of these applications, done in 2009 and motivated by the experimental results of Refs.~\cite{hertlein2008direct,gambassi2009critical}, consisted in obtaining the reversible aggregation, by temperature-induced CCFs, of colloidal particles immersed in a trimethylpyridine, water and heavy water critical mixture \cite{bonn2009direct}. Analyzing the experimental aggregations of the colloids, a simple theoretical model was proposed of the particle pair potential with the aim of explaining the observed aggregation line \cite{bonn2009direct}. However, this simple model turned out to be incomplete \cite{gambassi2010colloidal,bonn2010bonn,maciolek2018}. 
Afterwards, in 2012, it was shown in Ref.~\cite{faber2013controlling} that a reversible gas-liquid-solid transition of colloidal particles can be driven by CCFs in a trimethylpyridine, water and heavy water critical mixture. At temperatures sufficiently far from $T_{\rm c}$, the colloids were uniformly suspended into the solvent, forming a dilute gas phase; upon increasing the temperature towards $T_{\rm c}$, they were exhibiting the diffusive motion characteristic of liquids and at temperatures very close to $T_{\rm c}$ the colloids aggregated, forming an ordered face-centered cubic lattice. For a recent review on the role played by CCFs in controlling the aggregation of colloids we refer the reader to Ref.~\cite{maciolek2018}. 

In addition, by modifying the shape and the chemical properties of the colloidal surfaces, it is possible to produce patchy particles, whose interaction, bonds  and spatial aggregation can be tuned by CCFs, giving rise to the self-assembly of complex structures \cite{nguyen2016critical}. 
Moreover, CCFs can be employed to assembly colloidal quantum dots (QDs) into dense superstructures, with optoelectronic properties which can be controlled by the same CCFs \cite{marino2016assembling}.

In the following, after providing a simplified overview of the theory of critical phenomena and CCFs, we focus on the investigation of CCFs using optical tweezers describing the main experiments that have been performed in the last decade, before concluding with some perspective.

\section{Theory overview}

In order to understand better the nature of CCFs, we summarise here the basic facts of the theory of critical phenomena and describe briefly one of the physical systems which has been used to study them, i.e., the binary liquid mixture of water and 2,6 lutidine at its critical composition \cite{grattoni1993lower}. For a comprehensive review on critical phenomena, we refer the reader to Ref.~\cite{pellissetto2002}, while an overview of the theory of CCFs can be found in Refs.~\cite{krech1999,criticalphenomenabook,gambassi2009casimir,gambassi2011critical,maciolek2018}. Here, we follow a descriptive approach, reporting only the predictions of this theory.

\begin{figure}[h!]
	\begin{center}
	\includegraphics[width=12cm]{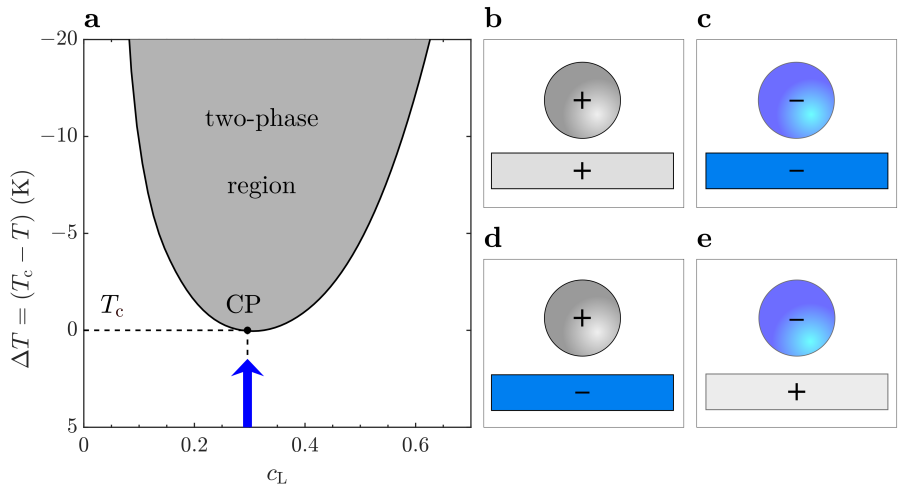}
	\caption{
	{\bf Phase diagram of the water$-$2,6 lutidine mixture, and schematic of the possible combinations of boundary conditions for a spherical particle and a flat surface.} (a)
	 Phase diagram of the water$-$2,6 lutidine mixture, drawn on the basis of the data of Ref.~\cite{grattoni1993lower}, featuring a lower critical point (CP) at the bottom of the coexistence line (thick solid line). Measurements of the potentials of the forces acting on the particle are typically performed at the critical lutidine mass fraction $c_{\rm L}^{c}=0.286$, while the temperature $T$ is gradually increased towards its critical value $T_{\rm c} = 307\, {\rm K}$, as indicated by the arrow.  
	 (b-e) Schematic illustration of the possible combinations of boundary conditions in a system constituted by a flat surface and a spherical particle immersed in the liquid mixture. Hydrophilic surfaces are indicated in gray and hydrophobic surfaces in blue. (b) The combination of two hydrophobic surfaces is referred to as symmetric $(++)$ boundary conditions, (c) while a combination of two hydrophilic surfaces is referred to as a symmetric $(--)$ boundary conditions. (d,e) Antisymmetric $(+-)$ (or $(-+)$) boundary conditions, instead, are realized by (d) a hydrophobic particle and a hydrophilic flat surface, or (e) a hydrophilic particle and a hydrophobic surface.
	}
	\label{figure:1:waterlutidine}
	\end{center}
\end{figure}

2,6-lutidine, or dimethylpyridine (C$_7$H$_9$N), is a natural organic compound which, at room temperatures lower than $34^{\circ}{\rm C}$ is miscible with water for a wide range of concentrations, forming an homogeneous, transparent solution. Upon gradually increasing the temperature, the solution eventually demixes: after observing turbidity and critical opalescence, the mixture separates in two distinct phases, one rich in water and the other in lutidine. 
For all values of lutidine concentration but one, this demixing phase transition is of the first order. However, at the concentration corresponding to the lower consolute point (CP), the phase transition becomes critical, i.e., of the second order. The phase diagram of  water$-$2,6 lutidine\cite{grattoni1993lower,gambassi2009casimir} in the vicinity of the lower critical point is represented schematically in Fig.~\ref{figure:1:waterlutidine}a.

The phase of a binary liquid mixture can be described by the order parameter $\delta c(\mathbf{r})$, i.e., the local deviation from the average concentration $ c_{0}$ of one of its two components: $\delta c(\mathbf{r}) = c(\mathbf{r}) - c_{0}$. 
When the mixture is homogeneous, the statistical average of the local concentration $\langle c(\mathbf{r}) \rangle$ is equal to the nominal concentration $c_{0}$ everywhere, while when the mixture is separated $\langle \delta c(\mathbf{r}) \rangle$ differs significantly from zero in the two separated phases. 

Upon approaching a phase transition, the thermally fluctuating order parameter is characterized by a certain spatial correlation length $\xi$, which is defined by the exponential decay of the correlation of the order parameter at different points $\mathbf{r}$ and $\mathbf{r}^{\prime}$: in fact, within a phase, $\langle \delta c(\mathbf{r}) \delta c(\mathbf{r}^{\prime}) \rangle - \langle \delta c(\mathbf{r}) \rangle \langle\delta c(\mathbf{r}^{\prime})\rangle$ decays as $\exp{\left( -\frac{|{\mathbf{r}-\mathbf{r}^{\prime} }|}{\xi}\right)}$ upon increasing $|{\mathbf{r}-\mathbf{r}^{\prime} }|$. $\xi$ depends  on both the temperature and the nominal concentration of the mixture. 
At first-order phase transitions, the correlation length $\xi$ remains finite. Upon approaching second-order phase transitions (i.e., approaching the critical point), instead, the correlation length $\xi$ increases without limits and becomes infinite when the critical temperature is reached\cite{gambassi2009casimir}. The exponent characterising the divergence of the correlation length  as a function of the distance from the critical temperature (at the critical concentration)  is referred to as the critical exponent $\nu$\ (see further below), which is the same for all the physical systems sharing the same gross features (symmetries, spatial dimensionality, range of interactions, etc.), i.e., belonging to the same universality class. In this sense, $\nu$ is a universal quantity\cite{pellissetto2002}. 

The mixture of water and 2,6-lutidine belongs to the Ising model universality class\cite{pellissetto2002}. This mixture has been extensively studied for its application in colloidal assembly and it presents several advantages: as mentioned above, it has a lower critical point at about $34^{\circ}{\rm C}$, which is very close to room temperature, it is non-toxic, and of relatively easy preparation. All these features make this mixture an ideal system to study critical phenomena and to observe CCFs in experiments.

As long as one is interested in determining universal quantities such as the critical exponent and the CCF (see below), one can choose any physical system belonging to the same universality class. As the Ising model describes the liquid-vapor transition in simple fluids, as well as the transitions occurring in multicomponent fluid mixtures, in uniaxial ferromagnets, and in micellar systems\cite{pellissetto2002}, there is a wide choice of systems belonging to the same universality class as the demixing transition of binary mixtures, which have been successfully employed in experiments. For example, a ternary mixture of trimethylpyridine, water and heavy water\cite{faber2013controlling} or micellar  solutions, such as C$_{12}$E$_5$ or C$_{12}$E$_8$ \cite{martinez2017energy,Buzzaccaro2010criticaldepletion}. 

When an object like a flat plate or a colloid is immersed in a critical water$-$2,6 lutidine mixture, its surface comes into contact with the two components of the mixture and, depending on its chemical properties, adsorbs preferentially one of them, {\em de facto} altering the composition of the mixture within a layer of typical thickness $\xi$ in contact with the surface. This imposes a constraint on the mixture local composition in the direct proximity of the immersed object. When a second plate is immersed in the solution at a distance $L$ from the first one, the local concentration fluctuations between the plates, characterised by their correlation length $\xi$, become confined and an effective interaction emerges between the two plates, with a typical range set by $\xi$.

The magnitude and the attractive or repulsive character of the CCF depend on the ratio $L/\xi$ and on the surface adsorption properties of the involved surfaces. 
The smaller the ratio $L/\xi$ is, the more relevant  the CCFs will be, because, correspondingly, the correlation length $\xi$ of the local concentration fluctuations becomes comparable with or larger than the typical length scale $L$ of the system and therefore one of the two surfaces falls within the range of distances at which the other affects, with its presence, the order parameter and its fluctuations. 
For a fixed ratio $L/\xi$, combinations of homogeneous surfaces differing in preferential adsorption  affect both the magnitude and the attractive or repulsive character of the force\cite{nellen2009tunability}: when both surfaces prefer the same component of the binary mixture, the interaction is found to be attractive, while when they prefer different components, the interaction is found to be repulsive. 

The different preferential adsorptions are  modelled theoretically via boundary conditions induced by the presence of a boundary field favoring one or the other component of the mixture (see Fig.~\ref{figure:1:waterlutidine}b-e: the schematics refers to a system of one particle and a flat surface, but it can be easily transposed to the case of the two flat surfaces discussed here). Two hydrophilic plates, both preferring water, constitute a system with symmetric boundary conditions, indicated by $(- -)$. Two hydrophobic plates, both preferring lutidine, constitute a system with symmetric boundary conditions,  indicated by $(+ +)$. A combination of a hydrophilic and a hydrophobic plate constitute a system with antysimmetric boundary conditions, indicated by $(- +)$.

The fact that the surface adsorption properties determine the attractive or repulsive character of the CCF has been predicted theoretically\cite{krech1992,krech1997} and confirmed experimentally by measuring the wetting layer thickness in different alkane/methanol critical mixtures \cite{Fukuto2005,rafai2007}. 
The different behaviour of the CCF for symmetric or antisymmetric boundary conditions has been demonstrated also in experiments with colloids suspended in a water--2,6-lutidine mixture in the proximity of either a homogeneous surface \cite{hertlein2008direct,gambassi2009critical} or a patterned surface with hydrophilic and hydrophobic regions\cite{soyka2008critical,trondle2011}. The direct quantitative measurement of the dependence of the CCF upon varying the boundary conditions has been done in  Ref.
~\cite{nellen2009tunability} using a surface with a gradient in its preferential adsorption properties: The particle was held near the surface on different regions, to test  different combinations of boundary conditions at constant temperature, and the critical Casimir interaction potential was extrapolated for each region, showing a dependence on the surface properties in agreement with the theoretical predictions.

The dependence on the boundary conditions allows one to tune the CCF between objects. However, the boundary conditions realized at each surface are determined either by the material constituting the surface  or by the surface treatment the material has been subject to and therefore it cannot be changed easily at will. 
The dependence of CCFs on the temperature provides, instead, a very effective way of modulating their strength. 
According to the theory of critical phenomena, as the temperature $T$ approaches the critical temperature $T_{\rm c}$ of a second-order phase transition from the homogeneous phase, the correlation length $\xi$ of the critical fluctuations increases, according to $\xi = \xi_{0} \left| \frac{T-T_{\rm c}}{T_{\rm c}} \right|^{-\nu}$.  $\xi_0$ is a  microscopic length which depends on the specific critical system in exam and also on the phase from which the critical point is approached, forming universal ratios\cite{pellissetto2002}, while $\nu$ is the universal critical exponent introduced above. In the presence of boundaries,  sufficiently close to the critical point, the correlation length becomes comparable to their distance and therefore the CCF becomes important for the objects.

The CCF acting on a surface can be derived from its critical Casimir potential $\Phi_{\rm C}$, which, in three spatial dimensions, is theoretically known to depend on a universal scaling function $\Theta(x)$ as 
\begin{equation}
\frac{\Phi_{\rm C}(L)}{S} = \frac{k_{\rm B}T}{L^2} \Theta\left(\frac{L}{\xi}\right),
\end{equation}
where $k_{\rm B}$ is the Boltzmann constant, $S$ the (large) area of the involved surface and $L$ the distance between the plates. 
In the presence of boundaries and sufficiently close to the critical point, $\Theta(x)$ depends on the universality class of the liquid medium undergoing the transition as well as on the geometry (which consequently affects the geometrical prefactor $S/L^2$ in the previous equation)  and on the boundary conditions of the system: once these gross features are determined, $\Theta$ depends solely on the ratio between the surface distance $L$ and the correlation length $\xi$. The scaling function does not depend on the temperature $T$ explicitly, but it does it  implicitly via $\xi$.
$\Theta(x)$ has been calculated theoretically for various universality classes and combinations of surfaces (in particular for planes, spheres, ellipsoids and cylinders) and boundary conditions (symmetric, antisymmetric) via Monte-Carlo simulations or within the mean-field approximation \cite{vasilyev2007,vasilyev2009,kondrat2009ellipsoidplane,hasenbusch2009,hasenbusch2010}. Similarly, the scaling function has been determined in more complex geometries \cite{trondle2008,hasenbusch2013,labbelaurent2014}, for Janus  \cite{labbelaurent2016,squarcini2020} or patchy \cite{farahmand2020} particles, and other chemically or topographically  patterned surfaces \cite{gambassi2011critical,sprenger2005,sprenger2006,trondle2010patterned,parisentoldin2010,trondle2015crenellated}. These predictions have been used to interpret various experimental observations \cite{hucht2007,vasilyev2007,vasilyev2009,soyka2008critical,trondle2011}.

As generically expected for fluctuation-induced forces, many-body effects have been predicted for CCFs, e.g., considering the lateral forces which arise in the presence of two colloids near a flat surface\cite{mattos2013}. A first experimental evidence of this nonadditivity  of potentials in the case of three spherical particles in the bulk of a water$-$lutidine mixture has been provided in Ref.~\cite{paladugu2016nonadditivity}.

\section{Investigation of critical Casimir forces by optical tweezers}

In this section, we will present in a chronological order a selection of experiments in which optical tweezers have been employed to measure CCFs as a fundamental part of the experimental setup\cite{hertlein2008direct,nellen2009tunability,martinez2017energy,paladugu2016nonadditivity,magazzu2019controlling}.
For a complete discussion of the additional technical details, concerning the structure and motivations of the various experimental setups, we refer the reader to the original articles, while  here we focus primarily on the concept of these experiments and on the role played by optical tweezers in the measurement.

\subsection{Direct Measurement of CCFs acting on a particle close to a flat surface}

\begin{figure}[b!]
	\begin{center}
    \includegraphics [width=12cm]{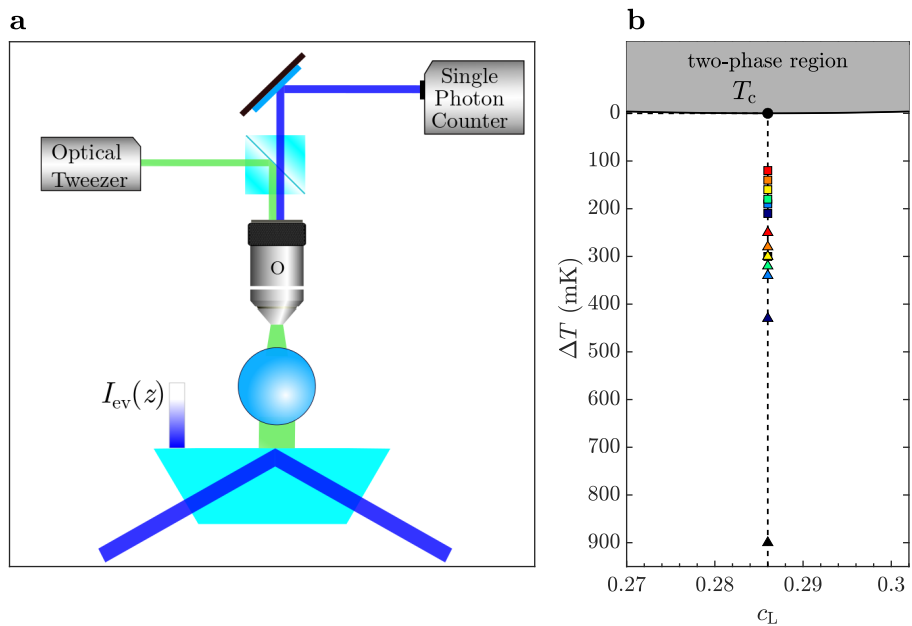}
	\caption{
	{\bf Scheme of the TIRM setup for the direct measurement of CCFs.}
    (a) TIRM setup, where an optical tweezers is used to hold a particle, immersed in a water--2,6-lutidine critical binary liquid mixture, at a predetermined position on the flat surface of a prism. The light beam propagating inside the prism is totally reflected by the surface, generating an evanescent wave of intensity $I_{\rm ev}(z)$ at a distance $z$ from the surface, which is scattered by the colloid. The scattered light is collected by and objective (O) and sent to a single-photon counter. The data acquired by the single-photon counter allow one to infer $z$, the distance of the particle from the flat surface, by inverting the known relation between $I_{\rm ev}$ and $z$.
	(b) Location, in the phase diagram of the water--2,6-lutidine mixture with lutidine mass concentration $c_L$, of the points at which the experiments in Ref.~\cite{hertlein2008direct} were done. The squares indicate the measurements in the symmetric $(--)$ case, corresponding to hydrophilic particle and hydrophilic substrate, while the triangles indicate the measurements in the antisymmetric $(+-)$ case, corresponding to hydrophobic particle and hydrophilic substrate. 
	Data  from Ref.~\cite{hertlein2008direct}.
	}
	\label{figure:2:setupHertlein}
	\end{center}
\end{figure}

\begin{figure}[h!]
	\begin{center}
    \includegraphics [width=12cm]{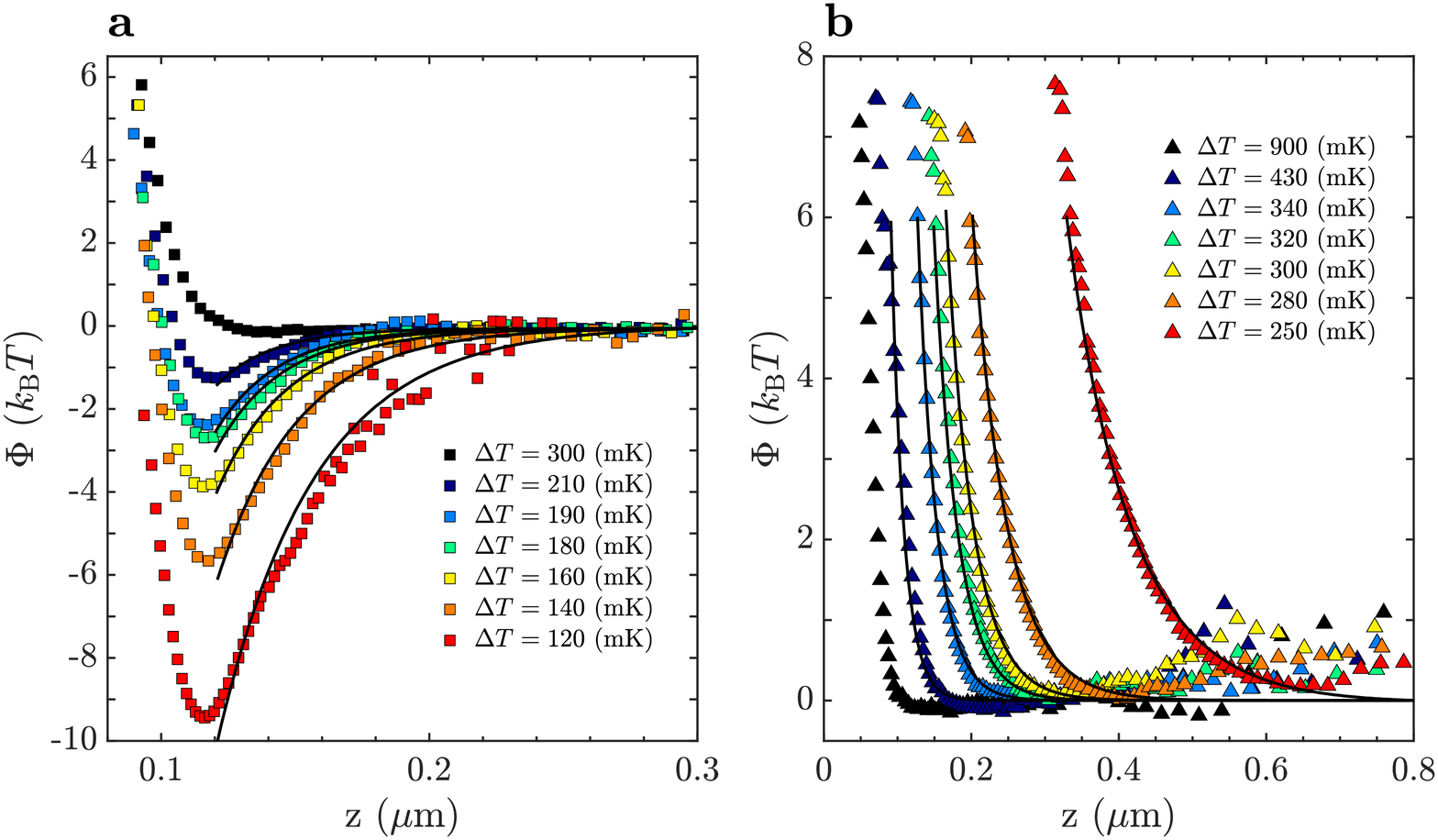}
	\caption{
	{\bf Direct measurement of CCFs.}
	 Potentials $\Phi(z)$ in units of $k_{\rm B}T$ as a function of the surface-to-surface distance $z$ between the particle and the flat surface. The experimental measurements at the various temperatures are represented by symbols of different shape and color. The theoretical prediction for each temperature is represented by the solid  black curves which are drawn only within the region in which van der Waals forces are negligible.
	 (a) Symmetric $(--)$ case: hydrophilic particle on hydrophilic substrate.
	 (b) Antisymmetric $(+-)$ case: hydrophobic particle on hydrophilic substrate.  Data  from Ref.~\cite{hertlein2008direct}.
	}
	\label{figure:3:directmeas}
	\end{center}
\end{figure}

The first direct measurement of the CCF between a spherical colloid and a planar boundary was performed in 2008\cite{hertlein2008direct}. 
A particle suspended in the water--2,6-lutidine mixture discussed above, at the critical composition, was held by optical tweezers at a certain position $z_0$ in the proximity of a planar boundary, as shown in Fig.~\ref{figure:2:setupHertlein}a. 
Although optically trapped, the particle fluctuates around its equilibrium position in the trap, because of the thermal agitation of the surrounding fluid. 
The total force acting on the particle is determined by the interplay of the optical forces and the other forces at play in the system, i.e., gravity, buoyancy, critical Casimir force, and, sufficiently close to the surface, also van der Waals and electrostatic interactions. 
The potential of the total force acting on the particle is obtained by monitoring its vertical motion and the statistics of the associated thermal fluctuations: this, in fact, is related to the potential energy in the explored space.
The vertical position of the particle with respect to the flat boundary was determined via TIRM, a technique which exploits the evanescent field created by the total internal reflection of a light beam: the evanescent wave is scattered by the colloid capturing it and by measuring the intensity of the scattered light it is possible to determine the position of the colloid \cite{hertlein2008direct}. 
Experimental acquisitions were done at the various temperatures indicated in the phase diagram in Fig.~\ref{figure:2:setupHertlein}b, in order to reconstruct the potential of the total forces acting on the particle and, after removing the optical and gravitational contributions, it was possible to obtain the interaction potential between the particle and the surface in each case. 

The results of the experiments are shown in Fig.~\ref{figure:3:directmeas}: the potential of the total force acting on the particle was measured for both symmetric and antisymmetric boundary conditions, for all inequivalent combinations (hydrophilic particle on hydrophilic substrate, hydrophilic particle on hydrophobic substrate, and hydrophobic particle on hydrophilic substrate) and the gravitational component could be easily subtracted from the raw data, obtaining the potential $\Phi$. In Fig.~\ref{figure:3:directmeas} this potential is reported as a function of the distance $z$ from the surface only for the case of an hydrophilic colloidal particle close to an hydrophilic substrate (Fig.~\ref{figure:3:directmeas}a)  and of an  hydrophobic particle close to the same substrate (Fig.~\ref{figure:3:directmeas}b); the data are taken from Ref.~\cite{hertlein2008direct}.
The experimental interaction potentials, represented by the various color-coded symbols corresponding to different temperatures $\Delta T$ according to the code used in Fig.~\ref{figure:2:setupHertlein}b, turned out to be in agreement with the theoretical predicted potentials\cite{gambassi2009critical}, represented by black solid lines, within the rage of values of the distance $z$ where the van der Waals forces and colloidal electrostatic interaction were negligible. 

Figure~\ref{figure:3:directmeas}a corresponds to the case of a suspended silica particle (hydrophilic surface) and it shows that the potential becomes increasingly more attractive upon decreasing $\Delta T$, i.e., upon approaching the critical temperature. Figure~\ref{figure:3:directmeas}b, instead, corresponds to the case of a suspended melamine particle (hydrophobic surface) and the potential becomes increasingly more repulsive upon approaching the critical point. From the potential $\Phi(z)$ within the range of distances at which only the critical Casimir potential contributes, the critical Casimir force $F(z)$ can be obtained straightforwardly, i.e., by calculating its gradient: $F(z) = -\frac{{\rm d}\Phi(z)}{{\rm d}z}$ .

\subsection{Tunability of CCF by boundary conditions}

\begin{figure}[b!]
	\begin{center}
	\includegraphics[width=12cm]{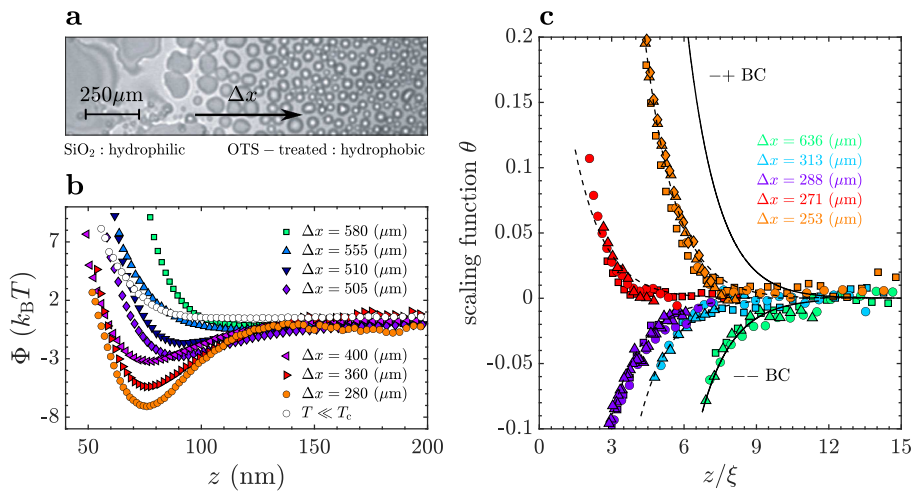}
	\caption{
	{\bf Tunability of CCFs by boundary conditions.}
	 (a) The different kind of surface adsorption properties of the slide are revealed by the varying wetting pattern formed by water on its surface. On the left of the image, the surface is hydrophilic, corresponding to a large contact angle, while it becomes increasingly less  hydrophilic, corresponding to a small contact angle, moving towards the center. On the right of the image, instead, the surface is hydrophobic, while it becomes increasingly less hydrophobic moving towards the center.
	 (b) Potential of the force on the colloidal particle, consisting of an electrostatic repulsion at short distances (revealed by the curve for $T \ll T_{\rm c}$) and of the critical Casimir potential $\Phi_{\rm C}(z)$, obtained for a fixed value of  $\Delta T$ but various boundary conditions, determined by the position $\Delta x$ along the slide of the experimental acquisitions.
	 (c) Experimental scaling functions $\theta = \frac{\Phi_{\rm C}(z)}{k_{\rm B}T} \times \frac{z}{R}$, being $R$ the radius of the colloid. Symbols with the same color but different shapes correspond to measurements taken at different temperatures which, after scaling, collapse onto a single curve for each value of $\Delta x$. The solid curves correspond to the theoretical predictions for perfectly symmetric boundary conditions ($--$ BC) and for perfectly antisymmetric boundary conditions ($-+$ BC). Data  from Ref. \cite{nellen2009tunability}.
	}
	\label{figure:4:tunability}
	\end{center}
\end{figure}

The experiment done in 2009 on the tunability of CCF by boundary conditions\cite{nellen2009tunability} is realised with the same experimental technique as in Ref.~\cite{hertlein2008direct}, i.e., TIRM,  with the difference that the planar surface above which the particle is suspended is a special substrate with a gradient in its preferential adsorption properties for the mixture components, as shown in Fig.~\ref{figure:4:tunability}a. The particle is held by optical tweezers at various lateral positions along the surface, thus experiencing the critical Casimir interaction with a substrate that goes from hydrophilic (on the left of Fig.~\ref{figure:4:tunability}a) to hydrophobic (on the right of Fig.~\ref{figure:4:tunability}a).

In Fig.~\ref{figure:4:tunability}b, the experimental potentials, which include the critical Casimir contributions and the electrostatic repulsion, measured at a given $\Delta T$ and different points along the glass slide are plotted. This is the demonstration that, for a certain temperature $\Delta T$ and a fixed surface-to-surface distance $z$, a change in the boundary conditions results in a change of the character of the force from attractive to repulsive, and therefore they can be used to  tune the force magnitude over a wide range. This can be seen by considering the potential at a certain distance  $z$   (for example, $z=80\,{\rm nm}$) from the substrate and by considering, the slope of the tangent line to the potentials (i.e., minus the force): the slope takes both positive and negative values upon varying the position $\Delta x$ along the substrate. This is the first main important result of Ref.~\cite{nellen2009tunability}.

In Fig.~\ref{figure:4:tunability}c, we show another important observation of the experiment, i.e., the emergence of an  experimental scaling function $\theta(z/\xi)$. This is obtained by plotting the critical Casimir contribution $\Phi_{\rm C}(z)$ to the potential per $k_{\rm B}T$, multiplied by $z/R$, where $R$ is the radius of the colloid,  as a function of the ratio $z/\xi$, $\xi$ being the correlation length corresponding to the various  temperatures of the mixture, and preferential adsorption properties of the region of plane above which the particle is trapped.

The theoretical predictions corresponding to perfectly symmetric $(--)$ and perfectly antisymmetric $(-+)$ boundary conditions, which physically correspond to having strong adsorption at the surfaces, are indicated by solid lines Fig.~\ref{figure:4:tunability}c. The black dashed lines represent a fit of the data with a shifted version of either of the two ideal cases, i.e., $(--)$ for identical adsorption properties, and $(-+)$ for complementary adsorption properties.
The experimentally determined scaling functions are color coded for the different adsorption properties, i.e., different position $\Delta x$ along the slide for the data acquisitions. For each position, data at different temperatures are plotted, and are indicated by symbols of different shape, but with the same color. Interestingly enough, while the experimental data indicate the emergence of the well-defined scaling function reported in Fig.  \ref{figure:4:tunability}c, at least within the range of parameters explored in the experiment, the theoretical analysis shows that scaling should be observed only after accounting for an additional scaling variable \cite{mohry2010crossover,abraham2010,vasilyev2011,hasenbusch2011}. 
This additional scaling variable depends on the effective strength of the critical adsorption (theoretically modeled by a surface ordering field $h_1$ which favors one or the other component of the liquid mixture at the boundary) rescaled, e.g., by a suitable power of the correlation length $\xi$, see, e.g., Ref.~\cite{vasilyev2011}  for additional details.  Formally, the $(-,\pm)$ boundary conditions correspond to having $|h_1| \to \infty$ with different or the same sign at the two confining boundaries.  In the experiment of Ref.~\cite{nellen2009tunability},
complete agreement with the theoretical prediction is retrieved in the case of identical adsorption properties and strong adsorption preference, while, when the surface has a less pronounced preference for the same mixture component, the scaling function appears as being {\em shifted} towards the origin, in the region in which only CCFs are relevant. For the case of opposite preferential adsorption properties, the experimental scaling function appears to be consistent with a shifted version of the case with perfect antisymmetric  boundary conditions, always in the range where only CCF are relevant. 

It is worth noting that complete and consistent interpretation of the experimental data in terms of the available theoretical predictions (see Ref.~\cite{vasilyev2011} which refers, however, to a different geometry) is still lacking, primarily due to the difficulty in having a quantitative experimental characterization of the local adsorption preferences of the substrate. 

\subsection{Energy transfer between colloids via critical interactions.}

\begin{figure}[h!]
	\begin{center}
	\includegraphics[width=12cm]{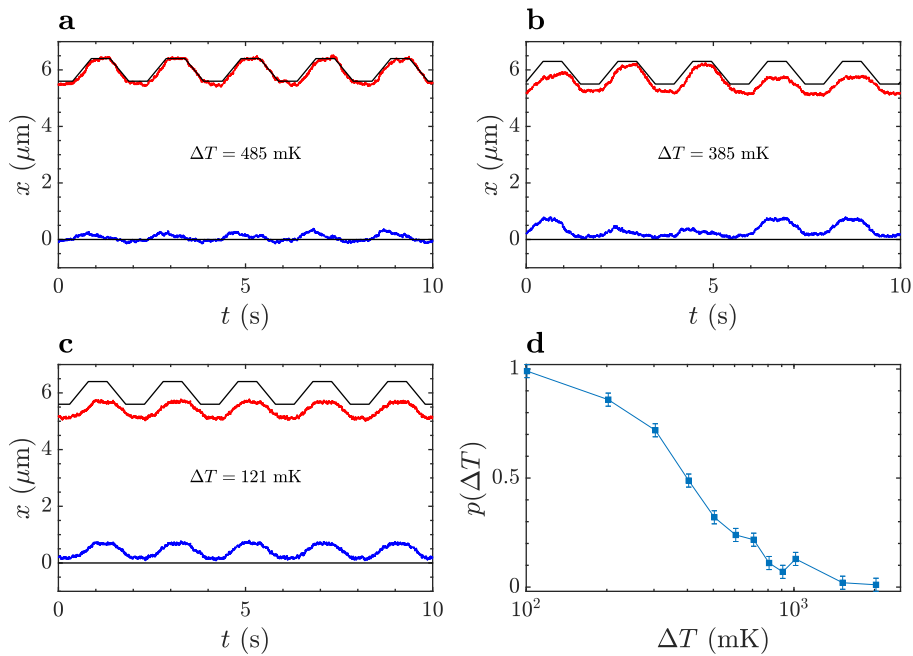}
	\caption{
	{\bf Particle synchronisation via CCFs. }
	{Trajectories of two trapped colloidal particles: particle 1 (blue) is subject to a spatially fixed optical trapping, while particle 2 (red) is subject to an optical trap whose center is moved according a specific protocol. The black solid lines represent the positions of the optical traps as functions of time. (a) No synchronization is observed between the particles for $\Delta T=485 \ \rm{mK}$ and, in fact, each of them, apart from Brownian fluctuations, follows the dynamics of the center of the corresponding trap. (b) Weak synchronization between the particles is observed for $\Delta T=385 \ \rm {mK}$. (c) Complete synchronization between the particles is achieved for $\Delta T=121 \ \rm {mK}$. 
    (d)	Probability $p (\Delta T)$ for the particles to be in a synchronous state.} Data  from Ref.~\cite{martinez2017energy}.
	}
	\label{figure:5:synchro}
	\end{center}
\end{figure}

\begin{figure}[h!]
	\begin{center}
	\includegraphics[width=12cm]{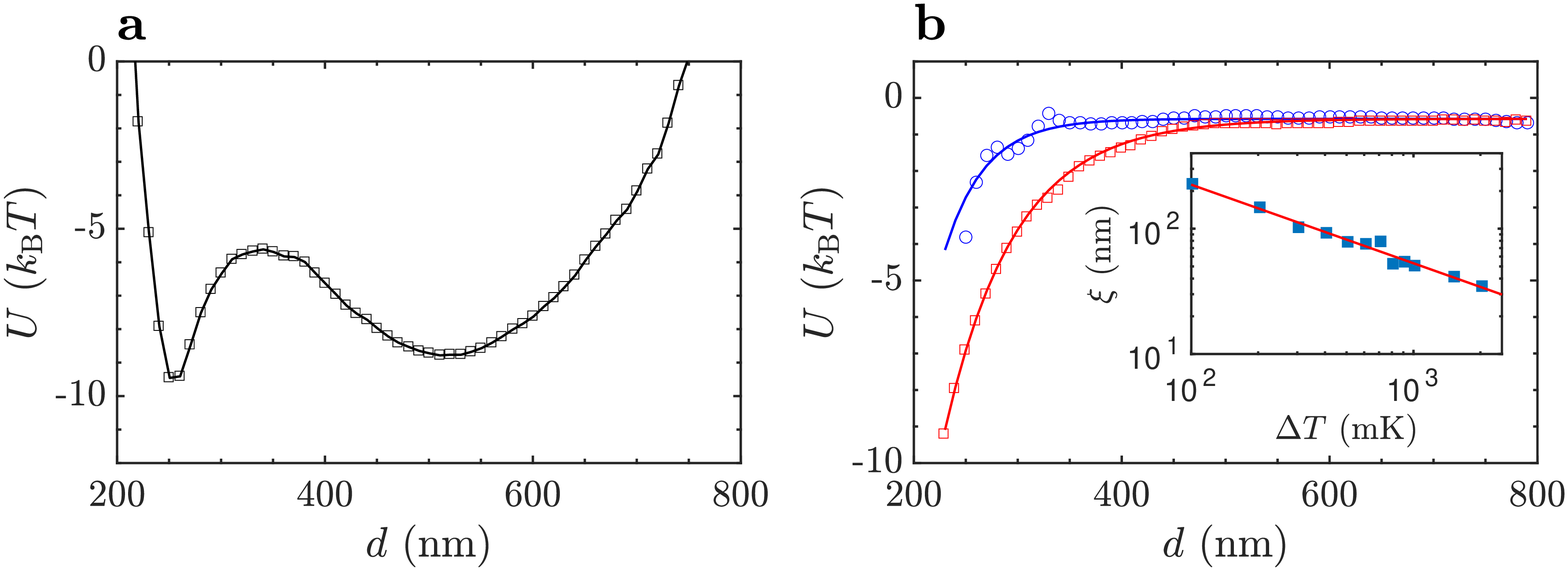}
	\caption{
	{\bf Interaction potentials between the particles. }
	(a) Total potential $U_{\rm total}(d)$ as function of the distance $d$ between the particles surfaces. (b) Critical Casimir potential for $\Delta T=2\ \rm K$ (blue circles) and $\Delta T=0.5\ \rm K$ (red squares), extracted from the experimental probability distribution of the relative distance between the colloids. Solid lines correspond to a fit using the theoretical prediction based on the Derjaguin approximation for the critical Casimir interaction, Using the value of the correlation length $\xi$ as a fitting parameter. Inset: values of $\xi$ obtained from the best fit of the experimental data with the theoretical predictions for various values of $\Delta T$, highlighting an algebraic dependence on $\Delta T$ with an exponent consistent with having $\nu = 0.63$, as expected for the universality class of the demixing transition\cite{pellissetto2002}. Data  from Ref.~\cite{martinez2017energy}.
	}
	\label{figure:6:potentials}
	\end{center}
\end{figure}

The effect of CCFs on the dynamics of particles is still a largely unexplored subject. Theoretical analyses based on numerical simulations \cite{furukawa2013} predict a rich dynamical behavior, including the emergence of retardation due to the motion of the colloids. The first experimental studies of dynamical aspects, in 2017, showed yet another interesting effect, i.e., that CCFs can induce synchronisation in the motion of a pair of colloids immersed in a binary liquid mixture close to the critical temperature \cite{martinez2017energy}, hence allowing the transfer of energy between colloids. 

In the experiment, a pair of colloids, identical in size (diameter $d \approx 5\, \upmu{\rm m}$) and surface properties, were held in two different optical traps generated by optical tweezers through an acousto-optical deflector, while immersed in the bulk of a micelle-solvent liquid binary mixture composed by C$_{12}$E$_5$ and water at $1.2\%$ of mass concentration. This solution is characterized by a lower critical point at $T_{\rm c}\approx 303.6 \ \rm K$, and a non-universal characteristic correlation length amplitude $\xi_0=1.4\ \rm{nm}$. Micelle-solvent critical solutions belong to the Ising universality class\cite{pellissetto2002}, and therefore the critical exponents and other universal features such as scaling functions of this system are the same as those of the water-2,6~lutidine critical binary liquid mixture used in the experiments mentioned above. A peculiarity of the C$_{12}$E$_5$-water micellar solution is that the characteristic length $\xi_0$ is approximately one order of magnitude larger than that of the $\xi_0$ of water-2,6 lutidine: as a consequence for the same value of the reduced temperature $\varepsilon = \frac{\Delta T}{T_{\rm c}}$, the former mixture has a correlation length $\xi$ which is approximately one order of magnitude larger than that of the latter, increasing effectively the spatial range of CCFs. 
The position of the colloids is recorded and analysed via digital video microscopy.

The spatial distance between the traps was chosen to have a surface-to-surface distance between the trapped particles of approximately $1 \ \upmu\rm m$. One trap was kept fixed at a certain position during the experiment. The other trap, instead, was made to oscillate continuously between two positions located along the line connecting the first trap with the initial position of the second trap, with an amplitude of approximately $0.4 \ \upmu\rm m$. In Fig.~\ref{figure:5:synchro}a-c, the position of the center of the oscillating trap as a function of time is represented by a black  solid line, while the center of the trap kept fixed is represented by the horizontal black solid line. 
Accordingly, apart from the Brownian motion, one expects the particle in the second trap to follow the position of the trap center. 
In the figure, the trajectory of the particle in the moving trap is represented by the set of red symbols, while the trajectory of the particle in the fixed trap by blue symbols.

When the temperature is far from its critical value, as in the case of Fig.~\ref{figure:5:synchro}a, the particle in the fixed trap fluctuates around a fixed position without being affected by the motion of the second particle, and there is no correlation between the displacements of the two particles. Upon gradually increasing the temperature towards $T_{\rm c}$, the motion of the particle in the fixed trap becomes increasingly synchronized with the motion of the particle in the moving trap, as shown in Figs.~\ref{figure:5:synchro}b and \ref{figure:5:synchro}c.  Similarly, the motion of the particle in the moving trap  turns out to be influenced by that of the particle in the fixed trap and therefore it no longer follows closely the solid line. 
This behaviour is due to the emergence of CCFs between the particles. While at low temperature the magnitude of the CCFs is negligible  at the typical distances between the particles, by increasing the temperature towards its critical value also the correlation length increases and, as in Fig.~\ref{figure:5:synchro}c, the CCFs become relevant within the whole range of interparticle distances dictated by the motion of the trap, making the particle in the fixed trap to follow that in the moving one. As anticipated, the average relative distance between the particles in Fig.~\ref{figure:5:synchro}a is different from that in Fig.~\ref{figure:5:synchro}c, despite the fact that the trap distance varies in time in the same way in the two cases. This is because, correspondingly, the CCF between the particles alter their equilibrium positions inside the two traps.

In order to quantify the degree of synchronization between these two motions, 
the probability $p (\Delta T)$ of the trajectories to be synchronized was calculated from the experimental data \cite{martinez2017energy}, and it is represented in Fig. \ref{figure:5:synchro}d. The probability that the particle trajectories are synchronized depends on the temperature of the sample, and it increases monotonically upon decreasing the distance form the critical point, due to the emergence of CCFs. 

The total potential $U_{\rm total}(d)$ between the particles was calculated from the extrapolated probability density function $\rho(d)$  of the surface-to-surface distance $d$, which is built from the relative trajectories of the particles. In Fig.~\ref{figure:6:potentials}a, an example of the total potential is reported for a certain value of $\Delta T$ at which the CCFs turn out to be relevant for the system, as witnessed by the presence of the characteristic dip  at a distance $d\simeq 250\,$nm in addition to the broader minimum at $d\simeq 520\,$nm determined by the trap.

In fact, the total potential is composed by three distinct contributions: the electrostatic repulsion at very short distances, the optical trapping potential (with the characteristic harmonic profile) at larger distances, and the attractive critical Casimir potential at intermediate distances. The theoretical expressions of these contributions are well-know for certain geometries and they can be used for others within the Derjaguin approximation, as  widely reported in literature, see, e.g., Refs. \cite{gambassi2010colloidal,martinez2017energy,paladugu2016nonadditivity,magazzu2019controlling,gambassi2011critical,gambassi2009critical,hertlein2008direct,gambassi2009casimir,trondle2011,vasilyev2007,vasilyev2009}. After subtracting from the total potential $U_{\rm total}(d)$ the electrostatic and optical potentials, one is left with the critical Casimir interaction potential. As done in the data analysis of previous experiments \cite{hertlein2008direct,nellen2009tunability}, the experimental values of the critical Casimir interaction were then fitted by the corresponding theoretical expression in order to determine the correlation length $\xi$ as fitting parameter for each experimental value of $\Delta T$, with the result reported in the inset of Fig. \ref{figure:6:potentials}b .
The agreement of the experimental data with the theoretical model (see in Fig. \ref{figure:6:potentials}b as far as both the critical Casimir force and the dependence on the correlation length on $\Delta T$ are concerned) confirmed the role of CCFs in the synchronization of the particles dynamics and showed that the critical Casimir effect within the present experiment is almost one order of magnitude stronger than that reported in the previous experiments, which explored typically smaller correlation lengths. In fact, in the  water-lutidine mixture used in Refs.~\cite{hertlein2008direct,nellen2009tunability,paladugu2016nonadditivity,magazzu2019controlling}, a distance $\Delta T \approx 500 {\ \rm mK}$ from the critical point corresponds to a correlation length $\xi \approx 10\ \rm{nm}$, while in the $\rm {C_{12}E_5}$-water critical solution employed in this experiment, the corresponding correlation length is $\xi \approx 70\ \rm{nm}$.

\subsection{CCFs between colloids in the bulk and experimental evidence of nonadditivity }

\begin{figure}[h!]
	\begin{center}
	\includegraphics[width=12cm]{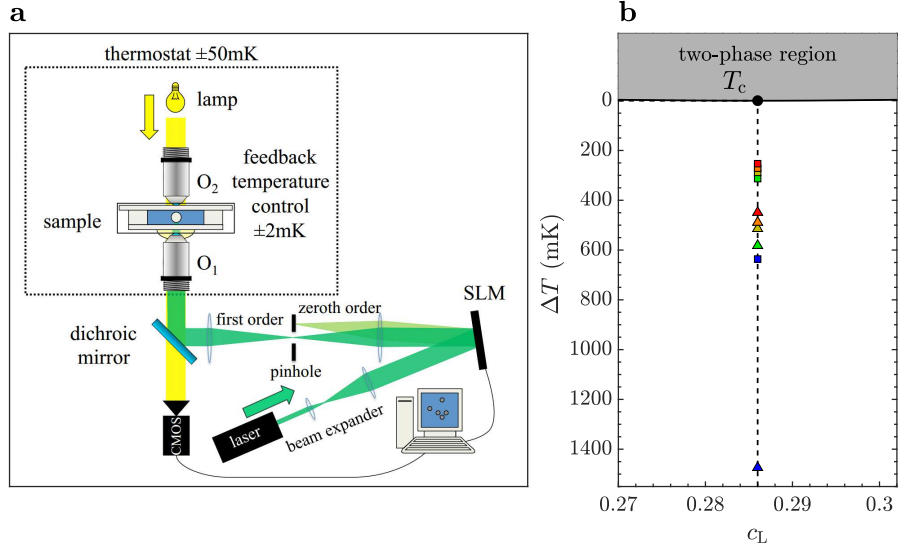}
	\caption{
	{\bf Setup for the direct measurement of CCFs between two particles in the bulk.}
	(a) The experimental setup consists of a holographic optical tweezers part, which generates two optical traps in the bulk of a water--2,6 lutidine critical binary liquid mixture, a digital video microscope, which tracks the positions of the colloidal particles, and a temperature control unit, for stabilizing the temperature of the focal region (dotted inset). For the detailed description of the experimental setup the reader is referred to Ref.~\cite{paladugu2016nonadditivity}.
	(b) Location, in the phase diagram of the water-lutidine mixture with lutidine mass concentration $c_L$ of the points at which the experimental acquisitions in Ref.~\cite{paladugu2016nonadditivity} were done. 
	The square symbols represent the experiments involving only hydrophilic particles (see Fig.~\ref{figure:8:hydrophilic}), while the triangles represent the experiments in which a third hydrophobic particle is later brought in the proximity of the first two hydrophilic particles (see Fig.~\ref{figure:9:hydrophobic}). 
	Figure and data reproduced from Ref.~\cite{paladugu2016nonadditivity}.
    }
	\label{figure:7:nonaddsetup}
	\end{center}
\end{figure}

\begin{figure}[h!]
	\begin{center}
	\includegraphics[width=12cm]{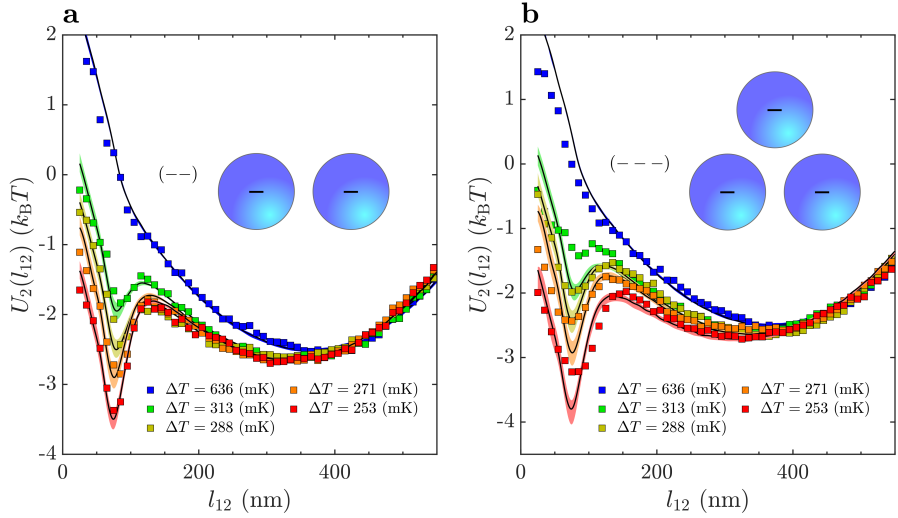}
	\caption{
	\textbf{CCFs between colloidal particles and their nonadditivity.}
	Total potentials of the forces acting on two hydrophilic particles which are optically trapped in the bulk of a water--2,6 lutidine critical mixture, (a) when far from other particles, and (b) in the proximity of a third hydrophilic particle. Symbols correspond to the experimental data, measured at the temperatures indicated in the legend. The solid lines correspond to the theoretical predictions for (a) the two-particle potential which accounts for electrostatic repulsion, optical forces and CCFs, and (b) the potential of the force acting on the same particles as before but now in the presence of a third particle, assuming, however, the validity of the linear superposition, i.e., the absence of nonadditivity. The insets in the panels are  schematic representations (out of scale) of the experimental traps configuration and of the positions of the trapped particles in each case. The effective potentials $U_{12}(l_{12})$ of the surface-to-surface distance $l_{12}$ between the two colloids are in units of $k_{\rm B}T$. The discrepancy between solid lines and data points in panel (b) demonstrate the nonadditivity of critical Casimir forces which becomes more pronounced upon approaching the critical point. Data  from Ref.~\cite{paladugu2016nonadditivity}.
	}
	\label{figure:8:hydrophilic}
	\end{center}
\end{figure}

\begin{figure}[h!]
	\begin{center}
	\includegraphics[width=12cm]{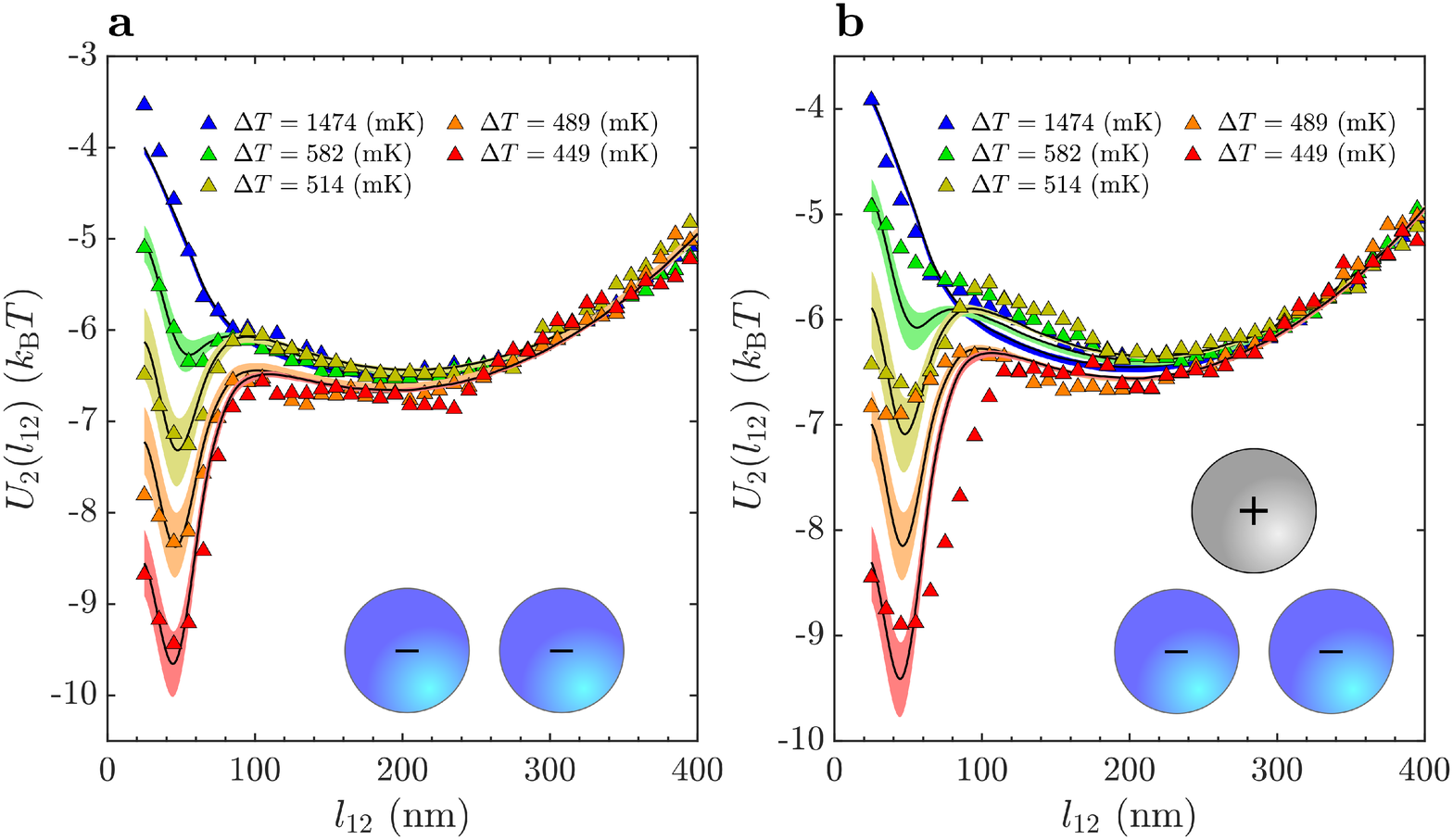}
	\caption{
	{\bf Nonadditivity of CCF with antisymmetric boundary conditions.}
	Total potentials of the forces acting on two hydrophilic particles which are optically trapped in the bulk of a water--2,6 lutidine liquid critical mixture, (a) when far from other particles, and (b) in the  proximity of a third hydrophobic particle. Symbols correspond to the experimental data, measured at the temperatures indicated in the legend. The solid lines correspond to the theoretical predictions for (a) the two-particle potential which accounts for electrostatic repulsion, optical forces and CCFs, and (b) the potential of the force acting on the same particles as before but now in the presence of a third particle, assuming, however, the validity of the linear superposition, i.e., the absence of nonadditivity.  The insets in the panels are  schematic representations (out of scale) of the experimental traps configuration and of the positions of the trapped particles in each case. The effective potentials $U_{12}(l_{12})$ of the surface-to-surface distance $l_{12}$ between the two colloids are in units of $k_{\rm B}T$. The discrepancy between solid lines and data points in panel (b) demonstrate the nonadditivity of critical Casimir forces, which becomes more pronounced upon approaching the critical point. Data  from  Ref.~\cite{paladugu2016nonadditivity}.
	}
	\label{figure:9:hydrophobic}
	\end{center}
\end{figure}

Critical Casimir forces belong to the large class of effective forces, which are theoretically expected to be nonadditive in the sense that, generally speaking, the effect obtained by a superposition of causes differs from the superposition of the effects of the single causes considered separately. The theoretical quantification of the nonadditivity of CCFs was first done, within a mean-field approximation,  in 2013\cite{mattos2013} for the configuration of two particles near a flat boundary. A measurement of nonadditivity in such a configuration, however, poses several experimental challenges, the major one being the impossibility of tracking independently two particles with TIRM.  
A viable technique for resolving the simultaneous trajectories of two particles is digital video microscopy, used, e.g., in Ref.~\cite{martinez2017energy}.  However, adding a second particle in the experimental setup of Ref.~\cite{hertlein2008direct} would rise the problem of measuring directly the CCFs between the nearby particles, which can be done in the bulk only. 
Because of this, in Ref.~\cite{paladugu2016nonadditivity} it was decided to use optical microscopy in order to measure first the CCFs between two particles in the bulk of a water--2,6-lutidine critical mixture, and then to add a third particle in their proximity to investigating the (non)additivity. 
In order to be able to hold particles in the bulk of a liquid, a configuration of optical traps was designed and generated via holographic optical tweezers, as explained in Ref.~\cite{paladugu2016nonadditivity}.

In order to test the (non)additivity of CCFs acting on the particles, it is necessary first to know and determine the  the mutual force between two isolated particles, so that one can later check whether the force on one of the two particles in the simultaneous presence of a third particle can actually be obtained as the vectorial sum of the mutual forces acting on the same particle and due to each of the two other particles taken separately. Instead of the force, one can conveniently consider the interaction potential, which is a scalar quantity, and proceed in the same logical way. The first step is to determine the critical Casimir interaction potential between two particles in the bulk of a water$-$2,6 lutidine critical mixture (see Fig.~\ref{figure:1:waterlutidine} for the phase diagram) for several values of the temperature $T$, or, equivalently, of the correlation length $\xi$. This is done by trapping two colloids (silica, diameter $d=2.06\pm 0.05 {\upmu}{\rm m}$) in the bulk of the critical binary liquid mixture, by using holographic optical tweezers, in a configuration where the traps are on 
a plane which is perpendicular to the optical axis of the acquisition camera, for a simpler evaluation of their center-to-center distance. The positions of the traps must be such that the particles are kept a few hundreds nanometers apart from each other (in the actual experiment, the trap distance was $L=2.38 \upmu {\rm m}$). 
Their trajectory in the object plane is measured with a frame rate of 200 fps. From the distribution of the center-to-center distance, the knowledge of the individual optical trap stiffnesses, of the colloidal electrostatic interaction, and of the theoretical prediction for the critical Casimir potential, one can fit the value of the actual correlation length and few other constrained parameters by comparing the experimental data with those of a numerical simulation, based on a Monte Carlo integration. 
The resulting effective potentials are reported in Figs.~\ref{figure:8:hydrophilic} and \ref{figure:9:hydrophobic}: symbols correspond to the experimental data, while the solid lines with a color shading indicate the theoretical predictions as explained below. 

Once the interaction potential between two particles is known and the few parameters entering the theoretical description of the interaction have been fit, a third particle is brought close to the first two. All the particles are tracked and the effective interaction potential of each couple of particles is determined by analysing the distribution of the corresponding center-to-center distance. This interaction potential, measured experimentally, is then compared with the interaction potential one would observe under the assumption that CCFs are additive.
If the experimental points turn out to be within the prediction of this additive model, with the corresponding uncertainties, then the CCF is additive in the explored experimental range. If the experimental points, instead, are outside the uncertainty range of the model in the region where CCFs are relevant, this provides a clear evidence of nonadditivity. In Figs.~\ref{figure:8:hydrophilic} and \ref{figure:9:hydrophobic} we show the results of the experiments  in the cases in which the third particle has the same preferential adsorption as the other two (i.e., they are all hydrophilic), as in Fig.~\ref{figure:8:hydrophilic}, or the opposite, as in Fig.~\ref{figure:9:hydrophobic}, corresponding to a third hydrophobic colloid.  Far from the critical temperature, no CCFs are relevant for the system, and no significant discrepancy is observed for the total interaction potential, as shown by the agreement between the solid line and the blue data points in panels (b) of Figs.~\ref{figure:8:hydrophilic} and \ref{figure:9:hydrophobic}. Upon increasing the temperature towards its critical value, a significant discrepancy between data points and solid lines emerges, particularly as far as the depth of the dip of the effective potential, due to the attractive CCF is concerned. The experimental dip is less pronounced than the theoretical prediction which assumes additivity, indicating that, in the presence of a third particle, the critical Casimir interaction potential is weaker than expected,  independently of the surface preferential adsorption of the third particle.  
This experiment provided the first clear demonstration of the generic nonadditivity of CCFs. 

\subsection{Effects of CCFs on the dynamics of colloidal particles}

\begin{figure}[t!]
\includegraphics [width=12cm]{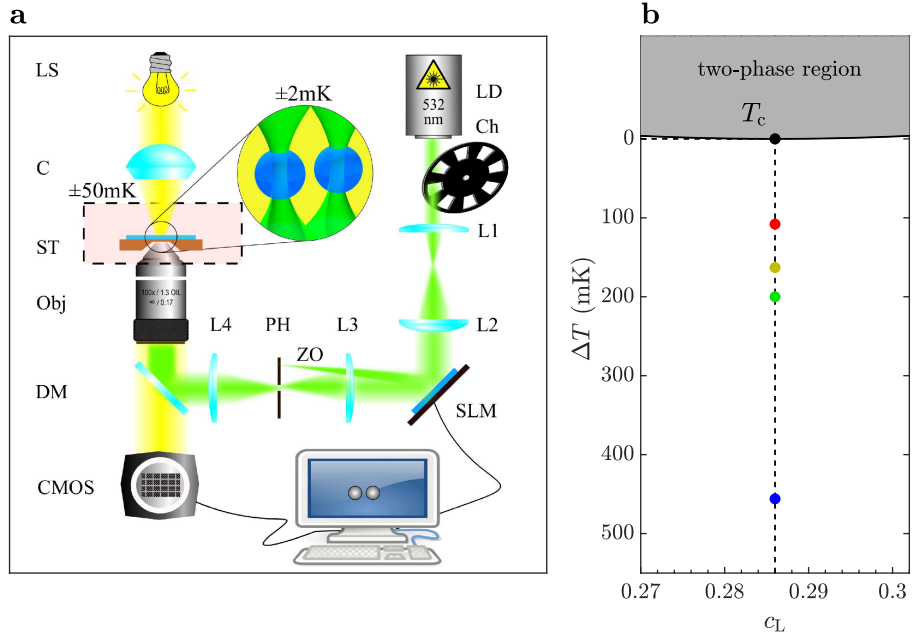}
\caption{
{\bf Blinking optical tweezers setup for the measurement of the effect of CCFs on the dynamics of colloidal particles.} 
(a) The experimental setup consists of a holographic optical tweezers part, which generates two optical traps in the bulk of a water--2,6 lutidine critical binary liquid mixture, a digital video microscope, which tracks the positions of the colloidal particles, and a temperature control unit, for stabilizing the temperature of the focal region (dashed inset). The setup is essentially the same as the one of Ref.~\cite{paladugu2016nonadditivity}, with the only addition of a chopper (Ch) which switches the traps on and off periodically. For additional details we refer the reader to Ref.~\cite{magazzu2019controlling}.
(b) Location, in the phase diagram of the water-lutidine mixture, with lutidine mass concentration $c_L$, of the points at which the experiments in Ref.~\cite{magazzu2019controlling} were done. Data  from Ref.~\cite{magazzu2019controlling}.}
\label{figure:10:blinkingsetup}
\end{figure}

\begin{figure}[t!] 
	\begin{center}
	\includegraphics[width=12cm]{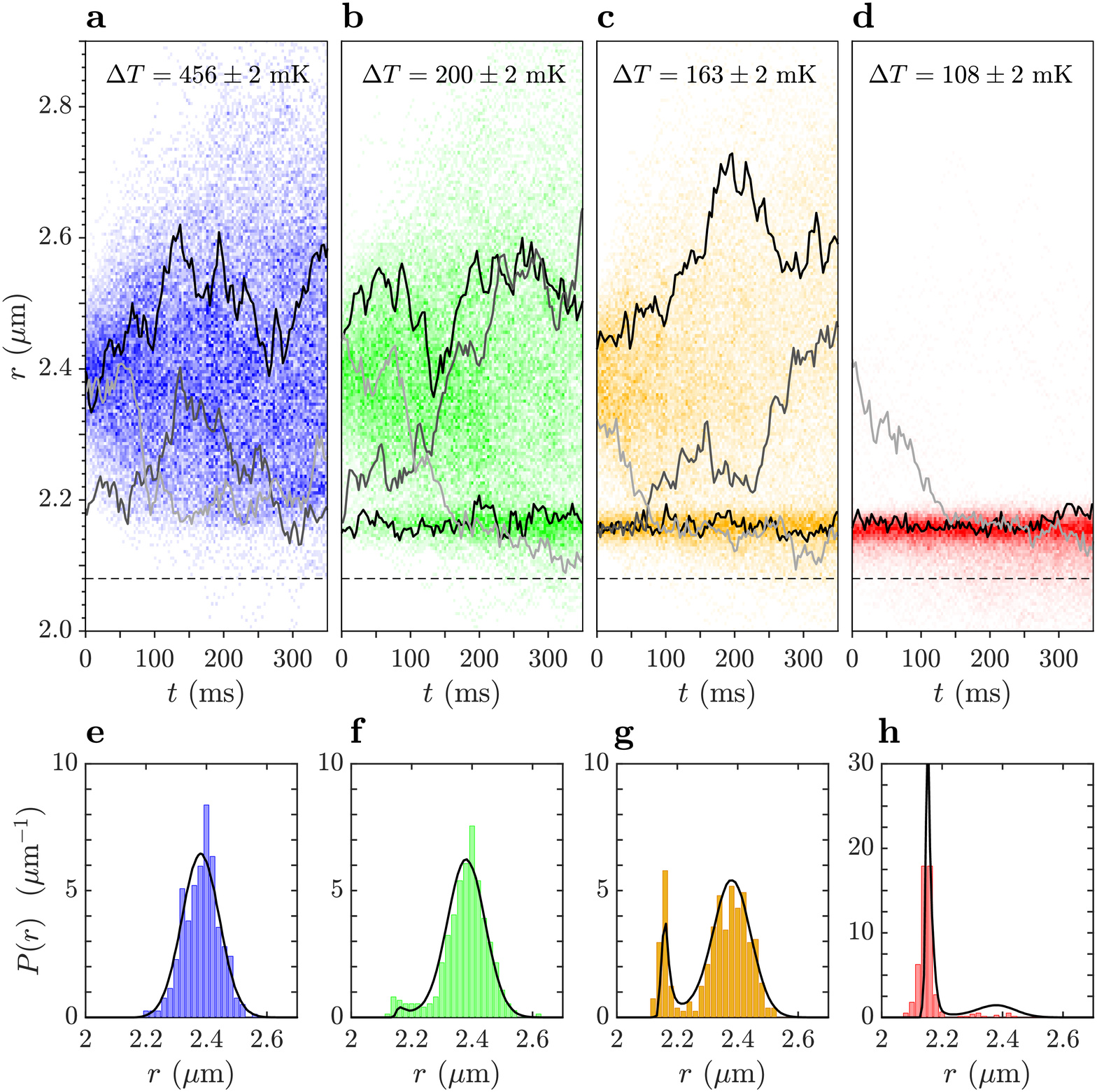}
	\caption{
	{\bf Time evolution of the inter-particle distance between unconfined colloids under the action of CCFs, and equilibrium distribution in the presence of optical confinement.} 
	(a-d) Time evolution of the probability density distribution, indicated by the intensity of the coloured background, of the inter-particles distance $r(t)$ obtained from 400 different trajectories after the optical traps are switched off at $t=0$ for decreasing values of $\Delta T$, indicated by the colorcode introduced in panel (b) of Fig. 10 and in each plot. The solid lines correspond to representative individual trajectories.
	The dashed horizontal line indicates the distance $r$ corresponding to the diameter $d$ of the colloids.
	(e-h) Equilibrium distribution $P_{\rm eq}(r)$ of the inter-particle distance $r(0)$ (i.e., when the optical tweezers are switched off) for two optically trapped colloids at the temperatures indicated by the colorcode and in the corresponding plots on the first row. Each histogram is obtained from 400 different experimental values. 
	The solid black lines are the theoretical  distribution of $r(0)$, obtained via Monte Carlo integration ($10^6$ samples) of two optically trapped particles subjected to the theoretical total potential consisting of the electrostatic, optical and critical Casimir force. Data  from Ref.~\cite{magazzu2019controlling}.
	}.
	
	\label{figure:11:blinkingtrajectories}
	\end{center}
\end{figure}

All the experimental studies of CCFs before 2019, with the sole  exception of Ref.~\cite{martinez2017energy}, focused on the time-independent properties of CCFs at equilibrium, leaving their effects on the dynamics of colloidal particles largely unexplored.
In order to shed light on this aspect of critical Casimir forces, in 2019, blinking optical tweezers were used to study the time-dependent effects of CCFs on a pair of colloidal particles \cite{magazzu2019controlling}, with the experimental setup described in Fig.~\ref{figure:10:blinkingsetup}a. Thanks to this technique, it was possible to observe the particle dynamics in the absence of confining optical potentials, revealing only their effective inter-particle interaction due to CCFs  \cite{magazzu2019controlling}.

In these experiments, two silica particles, with a diameter $d=2.06\pm 0.05 {\upmu}{\rm m}$ and having hydrophilic surfaces, were dispersed in a critical binary liquid mixture of water and 2,6-lutidine and held by two optical traps at a fixed center-to-center distance $r_0=2.4\ \rm {\upmu m}$ \cite{grier2003revolution,jones2015optical}. The value of $r_0$ was chosen such that the surface-to-surface distance between the particles was significantly larger than the range of the electrostatic repulsion between the particles, but comparable with the range of the CCFs, set by the correlation length of the critical fluctuations. By periodically chopping the laser beam, as sketched in Fig.~\ref{figure:10:blinkingsetup}a, the colloids are left free to move according to the Brownian motion,  under the sole effect of CCFs, i.e., in the absence of the optical potentials. Whenever the laser beam is restored, the resulting optical potentials bring the particles back to their initial positions, up to fluctuations, and the whole blinking process is repeated in order to acquire sufficient statistics for the analysis of the dynamics of the colloids. For the details of the experimental setup we refer the reader to Ref.~\cite{magazzu2019controlling}.

The blinking frequency, the length of the time intervals during which the optical traps were switched on, and the time interval of free evolution, during which the traps were switched off, were carefully chosen and adjusted in order to (i) minimize the effect of gravity, i.e., to avoid a significant displacement of the colloids along the vertical direction due to gravity, (ii) be sure that the colloidal particles return to their equilibrium condition after each interval of free evolution and right before the optical trap is again switched off, (iii) observe the free evolution for a sufficiently long time, and (iv) have enough statistics. The experimental data were acquired for the various values of $\Delta T$  reported in the phase diagram in Fig.~\ref{figure:10:blinkingsetup}b. The trajectories of the particles were measured by digital video microscopy.

Figure~\ref{figure:11:blinkingtrajectories}a-d shows the time evolution of the probability density of $r(t)$, the relative distance between the particle centers, obtained from 400 recorded particle trajectories, and the probability density at a certain time is represented by the corresponding intensity of the colored background for each value of $\Delta T$. Some trajectories are highlighted (solid lines) in order to illustrate a typical behaviour of $r(t)$.  
When the temperature of the liquid mixture is far from $T_{\rm c}$, like in panel (a), the particles diffuse freely from their initial positions (distributed according to the equilibrium distribution determined by the trap potentials) and no CCFs affect their dynamics. This is confirmed by the fact that the probability density of the interparticle distance is rather broad during its evolution compared to what is observed closer to criticality \cite{magazzu2019controlling}.
Upon decreasing the values of $\Delta T$, as in panels (b) and (c), CCFs become relevant. When the particles, in their erratic motion, come closer than a certain range controlled by the correlation length, the attractive CCFs bring and keep the colloids even closer,  as confirmed by  the emergence in the probability distribution of $r$ of a band of higher intensity, clearly visible in panels (b) and (c), located at $r\approx 2.16\ \rm {\upmu m}$, where the repulsive electrostatic forces and the attractive CCFs balance each other.
Upon further increasing the temperature towards $T_{\rm c}$, as in Fig.~\ref{figure:11:blinkingtrajectories}d, the CCFs become so strong to hinder the free diffusion of the particles and most of the values of $r(t)$ slightly fluctuate within a small interval around $r\approx 2.16\ \rm {\upmu m}$. 

Within the time windows during which the optical traps are switched on, the positions of the particles evolve under the action of a total potential $V_{\rm tot}$, reaching the equilibrium distribution $P_{\rm eq}(r)$ after a sufficiently long time, which is of the order of a few fractions of a second. This total potential is the sum of the optical potentials, of the colloidal electrostatic repulsive interaction, and of the CCFs. 

In order to check that the equilibrium distribution $P_{\rm eq}(r)$ is actually reached experimentally while the traps are switched on, the distribution of the relative distances of the starting points of the freely evolving trajectories was plotted, and compared with the predicted equilibrium distribution of two optically trapped particles interacting with colloidal electrostatic interaction and CCFs\cite{jones2015optical,paladugu2016nonadditivity,magazzu2019controlling}.
The results are reported in Fig.~\ref{figure:11:blinkingtrajectories}e-h. When $T$ is far from $T_{\rm c}$, as in panel (e), the experimental $P_{\rm eq}(r)$ is consistent with the equilibrium distribution of two optically trapped colloids interacting with the electrostatic interaction only. 
Upon increasing the temperature towards $T_{\rm c}$, i.e., upon reducing the values of $\Delta T$, a peak arise at $r\approx 2.16 \ \rm {\upmu m}$ on the left side of the distribution, as clearly and increasingly shown in panels (f), (g), and (h), consistently with the predicted equilibrium distribution of two optically trapped colloids interacting with the electrostatic interaction and the CCF\cite{paladugu2016nonadditivity,magazzu2019controlling}.
The agreement between the experimental and the simulated data, represented by the solid lines in panels (e)-(h), provides evidence of the validity of the theoretical description of the effects of CCFs on colloidal particles, which can be exploited in order to create a protocol for the application and fine tuning of CCFs in nanotechnology. 

\section{Outlook}

Optical tweezers and CCFs represent a powerful combination for the realisation of nanomachines, which are crucial for the development of nanotechnology, as recognized by the 2016 Nobel Prize in Chemistry awarded for the design and synthesis of molecular machines \cite{browne2010making}.

Optical tweezers proved to be crucial in the direct measurement of critical Casimir forces, in the investigation of  their dependence on temperature and boundary conditions, and in the study of how CCFs affect the dynamics of colloids, for controlling  the time scale of aggregation and the effective energy transfer between particles. The versatility of optical tweezers in both holding the colloids in place and manipulating them allowed the realisation of experiments that would have been  otherwise very difficult, if not impossible. 

One of the main challenges in nanotechnology is the presence and the relevance of thermal fluctuations at the nanoscale.
These fluctuations are often seen as a nuisance, making the behaviour of micro-engines less deterministic and therefore less predictable \cite{quinto2014microscopic}. In addition, in this regime, electromagnetic fluctuations are responsible for the emergent of strong van der Waals interactions which, in turn, produce stiction between the components of nanodevices and prevent them from working as desired.
A possible way to overcome these unwanted effects could be the generation of  CCFs {\em in situ} by controlling the critical fluctuations and their confinement between objects. This still proves to be challenging, but eventually could be achieved through the investigation of the spatio-temporal scale of critical fluctuations in situ, possibly resolving individual fluctuation modes. 
This could provide a way to allow an engineering of fluctuations, optimizing the performance of nanoengines, producing tunable and localized CCFs to prevent stiction in nano-devices and to assemble complex structures.

Moreover, the validated theoretical models of the critical Casimir interactions in various geometries and with possible inhomogeneous surfaces can be exploited to create a base protocol for the application and to fine-tune CCFs, opening novel routes towards their potential applications to nanotechnology, such as the dynamic and reversible self-assembly of nanodevices \cite{marino2016assembling,nguyen2017tuning}, the ability of driving nanomachines \cite{schmidt2018microscopic} and the synchronisation of complex mechanisms \cite{martinez2017energy} at the nano- and micro-meter scale. In addition, by patterning the particle surfaces or using anisotropic shapes it will be possible to create anisotropic critical Casimir interactions and torques inducing the three-dimensional self-assembly of complex structures  \cite{sacanna2010lock,faber2013controlling,labbelaurent2014,trondle2015crenellated,nguyen2016critical,labbelaurent2017,nguyen2017switching,farahmand2020,squarcini2020}.

%
%

%
%

\bibliography{biblio}


\end{document}